\documentclass[fleqn,a4paper,12pt]{article}
\usepackage{amssymb}
\usepackage{makeidx}
\usepackage{amsmath}
\usepackage{graphicx}
\usepackage{lscape}
\usepackage{a4}
\usepackage{epsfig}

 \newcommand{\newc}{\newcommand}
\newc{\beq}{\begin{equation}} \newc{\eeq}{\end{equation}}
\newc{\bea}{\begin{array}} \newc{\eea}{\end{array}}
\newc{\ri}{{\mathrm i}}
\newc{\bW}{{\mathbf W}}
\newc{\bR}{{\mathbf R}}
\newc{\bN}{{\mathbf N}}
\newc{\Psibar}{\overline\Psi}
\newc{\w}{{\bf w}}
\newc{\E}{{\mathbf{E}}}
\newc{\bp}{{\bf p}}
\newc{\ta}{\tilde a}
\newc{\bV}{{\bf V}}
\newc{\bfV}{{\bf V}}
\newc{\bfG}{{\bf G}}
\newc{\bx}{{\bf x}}
\newc{\bu}{{\bf u}}
\newc{\bP}{{\bf P}}
\newc{\bJ}{{\bf J}}
\newc{\bK}{{\bf K}}
\newc{\pd}{{\partial}}
\newc{\ti}{{\times}}
\newc{\bA}{{\bf A}}
\newc{\bE}{{\bf E}}
\newc{\bfn}{{\bf\nabla}}
\newc{\ho}{\hookrightarrow}
\newc{\ra}{\rightarrow}
\newc{\bv}{{\bf v}}
\newc{\bb}{{\bf b}}
\newc{\bc}{{\bf c}}
\newc{\bd}{{\bf d}}
\newc{\tbb}{\tilde{\bf b}}
\newc{\tbc}{\tilde{\bf c}}
\newc{\tbd}{\tilde{\bf d}}
\newc{\bz}{{\bf 0}}
\newc{\bun}{{\bf 1}}
\newc{\bL}{{\bf L}}
\newc{\bS}{{\bf S}}
\newc{\bB}{{\bf B}}
\newc{\br}{{\bf r}}
\newc{\sig}{{\mathbf\sigma}}
\newc{\eg}{{\it e.g.\ }}
\newc{\bpi}{{\mathbf\pi}}
\newc{\ie}{{\it i.e.\ }}
\newc{\etal}{{\it et al}}

\def\JPA#1#2#3#4{#2 #1 {\em J. Phys. A: Math. Gen.} {\bf #3} #4}

\def\AP#1#2#3#4{#2 #1 {\em Ann. Phys.} {\bf #3} #4}

\def\NCB#1#2#3#4{#2 #1 {\em Nuov. Cim.} {\bf #3 B} #4}

\def\IJTP#1#2#3#4{#2 #1 {\em Int. J. Theor. Phys.} {\bf #3} #4}
\def\CMP#1#2#3#4{#2 #1 {\em Comm. Math. Phys.} {\bf #3} #4}

\def\PNASUS#1#2#3#4{#2 #1 {\em Proc. Nat. Acad. Sci. U.S.} {\bf #3} #4}
\def\TMP#1#2#3#4{#2 #1 {\em Theor. Math. Phys.} {\bf #3} #4}

\def\TMP#1#2#3#4{#2 #1 {\em Theor. Math. Phys.} {\bf #3} #4}

\catcode`@=11 \long
\def\@caption#1[#2]#3{\par\addcontentsline{\csname
ext@#1\endcsname}{#1} {\protect\numberline{\csname
the#1\endcsname}{\ignorespaces #2}} \begingroup \small
\@parboxrestore \@makecaption{\csname fnum@#1\endcsname}
{\ignorespaces #3}\par \endgroup} \catcode`@=12

\begin{document} \begin{titlepage} \vskip 2cm
\begin{center} {\Large\bf  Galilean equations for massless fields} \footnote{E-mail: {\tt
niederle@fzu.cz},\ \ {\tt nikitin@imath.kiev.ua} } \vskip 3cm {\bf
J. Niederle$^a$, \\ and A.G. Nikitin$^b$ } \vskip 5pt {\sl
$^a$Institute of Physics of the Academy of Sciences of the Czech
Republic,\\ Na Slovance 2, 182 21 Prague, Czech Republic} \vskip 2pt
{\sl $^b$Institute of Mathematics, National Academy of Sciences of
Ukraine,\\ 3 Tereshchenkivs'ka Street, Kyiv-4, Ukraine, 01601\\}
\end{center}
\vskip .5cm \rm

\begin{abstract}Galilei-invariant equations for massless fields
are obtained  via contractions of relativistic wave equations.
It is shown that the
  collection
 of non-equivalent Galilei-invariant wave equations for massless fields with
 spin equal 1 and 0 is very reach and corresponds to various contractions of the representations of the
 Lorentz group to those of the Galilei one. It describes
 many physically consistent systems, e.g., those of electromagnetic fields in
 various media or Galilean Chern-Simon models. Finally, classification of
 all linear and a big group of non-linear
 Galilei-invariant equations for massless fields is presented.

\end{abstract}

\end{titlepage}

\setcounter{footnote}{0} \setcounter{page}{1}
\setcounter{section}{0}

\section{Introduction}

It was observed by Le Bellac and L\'evy-Leblond \cite{lebellac}
already in 1973
 that a non-relativistic limit of the Maxwell
equations is not unique. According to them (\cite{lebellac} p. 218),
the term "non-relativistic" means "in agreement with the the principle
of Galilei relativity". Moreover, they claimed that there exist two
Galilei invariant theories of electromagnetism which can be obtained
by appropriate limiting procedures starting with the Maxwell theory.

The  words "Galilean electromagnetism" themselves, introduced in
\cite{lebellac}, looked rather strange since it is pretty well known
that electromagnetic phenomena are in perfect accordance with the
Einstein relativity principle. On the other hand, physicists are
always interested when non-relativistic approximations are adequate,
which makes the results of the paper \cite{lebellac} quite popular.
The importance of such results are emphasized by the fact that a
correct definition of non-relativistic limit is by no means a simple
problem, in general, and in the case of massless fields in particular
(see, for example, \cite{Hol}).

Analyzing contents of the main impact journals in theoretical and
mathematical physics one finds  that an interest of research in
Galilean aspects of electrodynamics belongs to an evergreen
subject. Various approaches to the Galilei-invariant theories were
discussed briefly in \cite{NN1}.
Galilei electromagnetism was discussed in papers \cite{ijtp}-
\cite{Abreu} with using {\it reduction approach}
in which the Maxwell equations were generalized into a
(1+4)-dimensional Minkowski space and then reduced to the
Galilei-invariant equations. Such reduction is based on the fact that the
Galilei group is a subgroup of the generalized Poincar\'e group
(i.e., of the group of motions of the flat (1+4)-dimensional
Minkowski space). For reduction of representations of the group
$P(1,4)$ to those of the Galilei group see refs. \cite{FN1} and
\cite{FN2}.

In spite of that, the Galilei electromagnetism  contains constantly
many unsolved problems, for example, the question of a complete
description of all Galilean theories for vector and scalar massless
fields. In fact solution of this problem is the main issue of the
present paper.

In paper \cite{NN1} the indecomposable representations of the
homogeneous Galilei group  $HG(1,3)$ were derived, namely, all those
which when restricted to representations of the rotation subgroup of
the group $HG(1,3)$, are decomposed to the spin 0 and spin 1
representations. Moreover, their connection with representations of
the Lorentz group via the In\"on\"u-Wigner contractions
\cite{contraction} were studied in \cite{NN1} and \cite{NN2}. These
results allow to complete classification of the wave equations
describing  both massive and massless fields.

In the present paper we use our knowledge of indecomposable
representations of the homogeneous Galilei algebra $hg(1,3)$ from
\cite{NN1} and \cite{NN2} to derive the Galilei invariant equations
for vector and scalar massless fields. We shall show that, in
contrast to the corresponding relativistic equations for which there
are only two possibilities, namely, the Maxwell equations and the
equations for the longitudinal massless field, the number of
possible Galilei equations is very huge. The principal description
 of such
equations for vector and scalar fields is presented in Appendix where
a complete list of relative differential invariants is presented.
Among them there are
equations for fields with more  or less components  then in the
Maxwell equations.

These results can be clearly interpreted via representation and
contraction theories. As was proved in the papers \cite{NN1} and
\cite{NN2}, there is a large variety of possible contractions of
diverse representations of the Lorentz group to those of the Galilei
one and, consequently, many non-equivalent Galilei massless fields.
In the following sections we use these results from \cite{NN1} and
\cite{NN2} to describe connections of the relativistic and Galilei
theories for massless fields. These connections appear to be rather
non-trivial: some completely decoupled relativistic systems can be
contracted to coupled Galilei ones.

We pay a special attention to nonlinear Galilean systems for massless
fields. In particular Galilei-invariant Born-Infeld and
Chern-Simon systems are deduced. It is shown that in contrast to the
relativistic case there exist a Galilei-invariant Lagrangian in (1+3)-
dimensional space which includes a Chern-Simon term bilinear in
field components.

In Sections 2 and 3 we present some results of paper \cite{NN1}
concerning the classification of indecomposable representations of
the homogeneous Galilei group and the contractions of related
representations of the Lorentz group. These results are used in
Section 4 to classify all non-equivalent Galilei-invariant equations
of first order for vector and scalar massless fields. In Section 5
non-linear Galilei-invariant equations are derived including
Galilean Born-Infeld and Chern-Simon systems.
Finally, the results of
classification are summarized and discussed in Section 6.


\section{Indecomposable representations of the homogeneous Galilei group}

The Galilei group $G(1,3)$ consists of the following transformations
of temporal and spatial variables: \beq\label{11}\bea{l} t\to
t'=t+a,\\{\bf x}\to {\bf x}'={\bf R}{\bf x} +{\bf v}t+\bf b,\eea\eeq
where $a,{\bf b}$ and ${\bf v}$ are real parameters of time
translation, space translations and pure Galilei transformations
respectively, and $\bf R$ is a rotation matrix.

The homogeneous Galilei group $HG(1,3)$ is a subgroup of the group
$G(1,3)$ leaving invariant point $\textbf{x}=(0,0,0)$ at time $t=0$.
It is formed by space rotations and pure Galilei transformations,
i.e., by transformations (\ref{11}) with $a=0$ and ${\bf b}= 0$.

The Lie algebra $hg(1,3)$ of the homogeneous Galilei group includes
six basis elements, namely, three generators $S_a, a=1,2,3$ of the
rotation subgroup and three generators $\eta_a$ of the Galilei
boosts. These basis elements satisfy the following commutation
relations
 \beq\label{e3}\bea{l}
[S_a,S_b]=\ri\varepsilon_{abc}S_c,\\
    {[}\eta_a,S_b{]}=\ri\varepsilon_{abc}\eta_c\ \ \texttt{and}\ \
{[}\eta_a,\eta_b{]}=0.\eea\eeq

 All indecomposable representations\footnote
 {Let us remind that a representation of a group $\mathfrak{G}$ in a
 normalized vector space $\mathfrak{C}$ is
 {\it irreducible}
 if its carrier space $\mathfrak{C}$ does not includes subspaces
 invariant w.r.t. $\mathfrak{G}$. The representation is called
 {\it indecomposable} if $\mathfrak{C}$ does not include invariant
 subspaces $\mathfrak{C_I}$ which are orthogonal to
 $\mathfrak{C}\setminus \mathfrak{C_I}$ .
 Irreducible representations are
 indecomposable too but indecomposable representations can be reducible in
 the sense
 that their carrier spaces can include (non-orthogonal)
 invariant subspaces.} of $HG(1,3)$
which, when restricted to the  rotation subgroup, are decomposed to
direct sums of vector and scalar representations, were found in
\cite{NN1}. These representations (denoted as $D(m,n,\lambda)$) are
labeled by triplets of numbers: $n,m$ and $\lambda$. These numbers
take the values \beq\label{mn}-1\leq
(n-m)\leq2,\ \ \ n\leq3, \ \ \ \lambda=\left\{\bea{l}0\ \texttt{if}\ m=0,\\
1\ \texttt{if}\ m=2\ \texttt{or}\ n-m=2,\\ 0, 1\ \texttt{if}\ m=1,
n\neq3. \eea\right.\eeq

In accordance with (\ref{mn}) there exist ten non-equivalent
indecomposable representations $D(m,n,\lambda)$. Their carrier spaces
 can include
 three types of rotational scalars $A, B, C$  and five
 types of vectors ${\bf R}, {\bf U}, {\bf W}, {\bf K}, {\bf N}$ whose
 transformation laws with respect to
 the Galilei boost are \cite{NN2}:
 \beq\label{fin}\bea{l}
 A\to A'=A,\\B\to B'=B+{\bf v \cdot R},\\C\to C'=C+{\bf v\cdot U}+
 \frac12{\bf v}^2 A,\\
 {\bf R}\to{\bf R}'={\bf R},\\{\bf U}\to{\bf U}'={\bf U}+{\bf v}A,
 \\{\bf W}\to {\bf W}'={\bf W}+{\bf v}\times{\bf R},\\
 {\bf K}\to{\bf K}'={\bf K}+{\bf v}\times{\bf R}+{\bf v}A,\\
 {\bf N}\to{\bf N}'={\bf N}+{\bf v}\times{\bf W}+{\bf v}B+
 {\bf v}({\bf v\cdot\bf R})-\frac12{\bf v}^2{\bf R},\eea\eeq
 where ${\bf v}$ is a vector whose components are parameters of
 the Galilei boost, ${\bf v \cdot R}$ and
 ${\bf v}\times{\bf R}$ are scalar and vector products of vectors
 $\bf v$ and $\bf R$ respectively.

Carrier spaces of these indecomposable representations of the group
$HG(1,3)$ include such sets of scalars $A, B, C$
 and vectors ${\bf R}, {\bf U}, {\bf W}, {\bf K}, {\bf N}$
 which transform among themselves w.r.t. transformations
 (\ref{fin}) but cannot be split to a direct sum of invariant subspaces.
  There exist exactly ten such sets which are listed in the following equation:
  \beq\label{rep}\bea{rcl}\{A\}& \Longleftrightarrow&\ D(0,1,0),\\
  \{{\bf R}\}& \Longleftrightarrow&\ D(1,0,0),\\\{B,  {\bf R}  \}&\
  \Longleftrightarrow&\ D(1,1,0),\\\{A, {\bf U}\}&\ \Longleftrightarrow&\
D(1,1,1),\\
  \{A,{\bf U}, C\}&\ \Longleftrightarrow&\ D(1,2,1),\\
\{{\bf W},
 {\bf R}\}&\ \Longleftrightarrow&\ D(2,0,0), \\
 \{{\bf R, W}, B\}& \ \Longleftrightarrow&\ D(2,1,0),\\
  \{A, {\bf K}, {\bf R}\}&\ \Longleftrightarrow&\ D(2,1,1),\\
   \{A, B, {\bf K}, {\bf R}\}&\ \Longleftrightarrow&\ D(2,2,1),\\
   \{B, {\bf N},{\bf W}, {\bf R}\}&\ \Longleftrightarrow&\ D(3,1,1).\eea
 \eeq

Thus, in contrary to the relativistic case, where there are only
three Lorentz covariant quantities which transform as vectors or
scalars under rotations (i.e., a relativistic four-vector,
antisymmetric tensor of the second order and a scalar), there are
ten indecomposable sets of the Galilei vectors and scalars which we
have enumerated in equation (\ref{rep}).

\section{Contractions of representations of the Lorentz algebra}

It is well known that the Galilei algebra can be obtained from the
Poincar\'e one by a limiting procedure called "the In\"on\"u-Wigner
contraction" \cite{contraction}. Representations of these algebras
can also be connected by this kind of contraction. However, this
connection is more complicated for two reasons. First, contraction
of a non-trivial representation of the Lorentz algebra yields to the
representation of the homogeneous Galilei algebra in which
generators of the Galilei boosts are represented trivially, so that
to obtain a non-trivial representation it is necessary to apply in
addition a similarity transformation which depends on a contraction
parameter in a tricky way. Second, to obtain indecomposable
representations of $hg(1,3)$ it is necessary, in general, to start
with completely reducible representations of the Lie algebra of the
Lorentz group.

In paper \cite{NN1} representations of the Lorentz group which can
be contracted to representations
 $D(m,n,\lambda)$ of the Galilei group were found
 and the related contractions specified. Here we present only
 a part of the results from \cite{NN1} which will be used in what follows.

Let us begin with the representation $D(\frac12,\frac12)$ of the Lie
algebra $so(1,3)$ of the Lorentz group, whose carrier space is
formed by four-vectors. Basis of this representation is given by
$4\times4$ matrices of the following form:
\beq\label{repa}S_{ab}=\varepsilon_{abc}\left(\bea{cc}s_c&\bz_{3\times1}\\
\bz_{1\times3}&0\eea\right),\ \ S_{0a}=
\left(\bea{cc}\bz_{3\times3}&- k^\dag_a\\k_a&0\eea\right).\eeq Here
$s_a$ are
 matrices of spin one with the elements $(s_a)_{bc}=i
\varepsilon_{abc}$ and $k_a$ are $1\times3$ matrices of the form
\beq\label{k} k_1=\left (\ri, 0, 0\right),\qquad k_2=\left (0, \ri,
0\right), \qquad k_3=\left (0, 0, \ri\right).\eeq

The In\"on\"u-Wigner contraction consists of the transformation to a
new basis
\[S_{ab}\to S_{ab}, \ S_{0a}\to\varepsilon S_{0a}\]
followed by a similarity transformation of all basis elements
$S_{\mu\nu}\to S'_{\mu\nu}=VS_{\mu\nu}V^{-1}$ with a matrix $V$
depending on a contraction parameter $\varepsilon$. Moreover, $V$
 depends on $\varepsilon$ in such a way that all
transformed generators $S'_{ab}$ and $\varepsilon S'_{0a}$ are kept
non-trivial and non-singular when $\varepsilon\to 0$
\cite{contraction}.

There exist two matrices $V$ for representation (\ref{repa}), namely:
\beq\label{con1}V_1=\left(\bea{rc}\varepsilon
I_{3\times3}&\bz_{3\times1}\\
\bz_{1\times3}&1\eea\right)\ \  \texttt{and  }
V_2=\left(\bea{rc}
I_{3\times3}&\bz_{3\times1}\\
\bz_{1\times3}&\varepsilon\eea\right).\eeq

Using $V_1$ we obtain
\beq\label{con3}S'_{ab}=V_1S_{ab}V_1^{-1}=S_{ab},\ \
S'_{0a}=\varepsilon
V_1S_{0a}V_1^{-1}=\left(\bea{cc}\bz_{3\times3}&-\varepsilon^2k^\dag_a
\\k_a&0\eea\right).\eeq
Then, passing $\varepsilon$ to zero, we come to the following
matrices \beq\label{con4}S_a=\frac12\varepsilon_{abc}S_{bc}=
\left(\bea{cc}s_a&\bz_{3\times1}\\
\bz_{1\times3}&0\eea\right),\ \eta_a=\lim
S'_{0a}|_{\varepsilon\to0}=
\left(\bea{cc}\bz_{3\times3}&\bz_{3\times1}
\\k_a&0\eea\right).\eeq

Analogously, using matrix $V_2$ we obtain \beq\label{con5}S_a=
\left(\bea{cc}s_a&\bz_{3\times1}\\
\bz_{1\times3}&0\eea\right),\ \eta_a=
\left(\bea{cc}\bz_{3\times3}&-k^\dag_a
\\\bz_{1\times3}&0\eea\right).\eeq

Matrices (\ref{con4}) and (\ref{con5}) satisfy commutation relations
(\ref{e3}), i.e., they realize representations of the algebra
$hg(1,3)$. More precisely, they form generators of indecomposable
representations $D(1,1,0)$ and $D(1,1,1)$ of the homogeneous Galilei
group respectively. Indeed, denoting vectors from the related
representation spaces as
\[\Psi=\left(\bea{c}R_1\\R_2\\R_3\\B\eea\right) \ \ \texttt{for }
D(1,1,0)\ \ \texttt{and
}\tilde\Psi=\left(\bea{c}U_1\\U_2\\U_3\\A\eea\right) \ \ \texttt{for
} D(1,1,1)\] and using the transformation laws (\ref{fin}) for $A,\
B,\  {\bf R}=\texttt{column}(R_1,R_2,R_3) $ and $ \ {\bf
U}=\texttt{column}(U_1,U_2,U_3)$ we easily find the corresponding
Galilei boost generators $\eta_a$ in the forms (\ref{con4}) and
(\ref{con5}). As far as rotation generators $S_a$ are concerned
they are direct sums of matrices of spin one (which are responsible
for transformations of 3-vectors $\bf R$ and $\bf U$) and zero
matrices (which keep scalars $A$ and $B$ invariant).

To obtain the five-dimensional representation $D(1,2,1)$ we have
 to
start with a direct sum of the representations $D(\frac12,\frac12)$
and $ D(0,0)$ of the Lorentz group. The corresponding generators of
the algebra $so(1,3)$ have the form \beq\label{Smu}\hat
S_{\mu\nu}=\left(\bea{cc}S_{\mu\nu}&\cdot\\\cdot&0\eea\right),\eeq
where $S_{\mu\nu}$ are matrices (\ref{repa}) and the dots denote
zero matrices of appropriate dimensions. The matrix of the
corresponding
  similarity transformation can be written as:
\beq\label{mat}V_3=\left(\bea{lrl}
I_{3\times3}&0_{3\times1}&0_{3\times1}\\0_{1\times3}
&-\frac12\varepsilon&\frac12\varepsilon
\\0_{1\times3}&\varepsilon^{-1} &\varepsilon^{-1}\eea\right),\ \
V_3^{-1}=\left(\bea{lrl}
I_{3\times3}&0_{3\times1}&0_{3\times1}\\0_{1\times3}&-\varepsilon^{-1}
&\frac12\varepsilon
\\0_{1\times3}&\varepsilon^{-1} &\frac12\varepsilon\eea\right).\eeq

As a result we obtain the following basis elements of representation
$D(1,2,1)$ of the algebra $hg(1,3)$:
\beq\label{con7}S_a=\left(\bea{ccc}s_a&\bz_{1\times3}
&\bz_{1\times3}\\\bz_{3\times1}&0&0\\
    \bz_{3\times1}&0&0\eea\right),\ \ \
    \eta_a=\left(\bea{ccc}\bz_{3\times3}&k^\dag_a&\bz_{3\times1}\\
  \bz_{1\times3}&0&0\\
    k_a&0&0\eea\right).\eeq

    Matrices $\eta_a$ in (\ref{con7}) generate transformations
    (\ref{fin}) of the components of vector-function
    $\hat\Psi=\texttt{column}({\bf U}, A, C)$ so that the relation
    \[\hat\Psi\to\hat\Psi'=
    \exp(\ri{\mbox{\boldmath$\eta\cdot v$\unboldmath}})\hat\Psi\]
    being written componentwise, presents the
    transformation properties for ${\bf U}, A$ and C written  in
    (\ref{fin}).

    Considering the representation $D(1,0)\oplus D(0,1)$ of the
    Lorentz group whose generators are $6\times6$ matrices
\beq\label{repb}S_{ab}=\varepsilon_{abc}\left(\bea{cc}s_c&\bz_{3\times3}\\
\bz_{3\times3}&s_c\eea\right)\ \ \texttt{and  } S_{0a}=
\left(\bea{cc}\bz_{3\times3}&-
s_a\\s_a&\bz_{3\times3}\eea\right),\eeq
    we have shown in
     \cite{NN1} that it can be contracted only to one indecomposable
     representation of the $HG(1,3)$, namely, to the representation
     $D(2,0,0)$. The corresponding contraction
     matrix can be chosen in the following form \cite{NN1}
\beq\label{con55}V_4=\left(\bea{rc}\varepsilon
I_{3\times3}&\bz_{3\times3}\\
\bz_{3\times3}&I_{3\times3}\eea\right).\eeq

In the present paper we shall use also another contraction matrix:
\beq\label{con6}V_5=\left(\bea{cr}
I_{3\times3}&\bz_{3\times3}\\
\bz_{3\times3}&\varepsilon I_{3\times3}\eea\right).\eeq Matrices
$V_4$ and $V_5$ are unitary equivalent but can lead to different
results when applied to search for Galilei limits of relativistic
equations whose solutions form a carrier space of the representation
$D(1,0)\oplus D(0,1)$ but also include additional dependent
variables. It happens, e.g., when one considers Galilei limits of
the Maxwell equations with currents and charges.

\section{Galilei massless fields}

For constructions of Galilei massless equations it is possible to
use the same approach as in \cite{NN3} where equations for the
massive fields has been derived. However here we prefer to apply
another technique which consists in  contractions of the appropriate
relativistic wave equations.

\subsection{Galilei limits of the Maxwell equations}

According to the L\'evy-Leblond and Le Bellac  analysis from 1967
\cite{ll1967}, \cite{lebellac} (see also
\cite{levyleblond},\cite{levycarroll}) there are two Galilean limits
of the Maxwell equations.

In the so-called "magnetic" Galilean limit we receive pre-Maxwellian
electromagnetism. The corresponding equations for  magnetic field
$\bf H$ and electric field $\bf E$
read\beq\label{mag}\bea{l}\nabla\times {\bf E}_m-\frac{\partial {\bf
H}_m}{\partial t}=0, \ \ \nabla\cdot{\bf
E}_m=ej^0_m,\\
\nabla\times {\bf H}_m=e{\bf j}_m,\ \ \ \ \ \ \ \ \nabla\cdot{\bf
H}_m=0,\eea\eeq where $j=(j^0_m,{\bf j}_m)$ is an electric current
and $e$ denotes an electric charge.

Equations (\ref{mag}) are invariant with respect to the Galilei
transformations (\ref{11}) provided vectors ${\bf H}_m, \ {\bf E}_m$
and electric current $j$ cotransform as \beq\label{tran}\bea{l}{\bf
H}_m\to{\bf H}_m,\ {\bf E}_m \to {\bf E}_m-{\bf v}\times{\bf
H}_m,\\{\bf j}_m\to{\bf j}_m,\ \ \ \ j^0_m\to j^0_m+{\bf v}\cdot{\bf
j}_m.\eea\eeq

Introducing a Galilean vector-potential $A_m=(A^0,{\bf A})$ such
that \beq\label{pot}{\bf H}_m=\nabla\times{\bf A},\ \ {\bf
E}_m=\frac{\partial {\bf A}}{\partial t}-\nabla A^0\eeq we obtain
from (\ref{tran}) the following transformation laws for $A$: \beq
A^0\to A^0+{\bf v}\cdot{\bf A}, \ {\bf A}\to{\bf A}.\label{A^0}\eeq

The other Galilean limit of the Maxwell equations, i.e., the
"electric" one looks as \beq\label{el}\bea{l}\nabla\times {\bf
H}_e+\frac{\partial {\bf E}_e}{\partial
t}=e{\bf j}_e, \ \ \nabla\cdot{\bf E}_e=ej^4_e,\\
\nabla\times {\bf E}_e=0,\ \ \ \ \ \ \ \ \ \ \ \ \ \nabla\cdot{\bf
H}_e=0,\eea\eeq with the Galilean transformation lows of the
following form \beq\label{t}\bea{l}{\bf H}_e\to{\bf H}_e+{\bf
v}\times{\bf E}_e,\ \ {\bf E}_e\to{\bf E}_e,
\\{\bf j}_e\to{\bf j}_e+{\bf v}j^4_e,\ \ \ \ \ \ \ \ \ \ \
j^4_e\to j^4_e.\eea\eeq Vectors ${\bf H}_e$ and ${\bf E}_e$ can be
expressed via vector-potentials as \beq\label{po}{\bf
H}_e=\nabla\times{\bf A},\ \ \ {\bf E}_e=-\nabla A^4\eeq with the
corresponding Galilei transformations for the vector-potential: \beq
A^4\to A^4, \ {\bf A}\to{\bf A}+{\bf v}A^4.\label{A^4}\eeq

The Galilean limits of the Maxwell equations (found in \cite{lebellac},
\cite{ll1967})
  admit clear interpretation in the representation
theory. The thing is that there are exactly two non-equivalent
representations of the homogeneous Galilei group the carrier spaces
of which are four-vectors -- the representations $D(1,1,0)$ and
$D(1,1,1)$. In other words, there are exactly two non-equivalent
Galilei transformations for four-vector-potentials and currents,
which are given explicitly in equations (\ref{A^0}), (\ref{tran})
and (\ref{A^4}), (\ref{t}). Equations for massless fields invariant
with respect to these transformations are written in (\ref{mag}) and
(\ref{el}).

Let us note that both representations, i.e.,  $D(1,1,0)$ and
$D(1,1,1)$, can be obtained via contractions of the representation
$D(1/2,1/2)$ of the Lorentz group whose carrier space is formed by
relativistic four-vectors. The related contraction matrices are
written explicitly in (\ref{con1}). Each of these contractions
generates a Galilei limit of the Maxwell equations either in the
form (\ref{mag}) or (\ref{el}). In Section 4 we shall obtain
equations (\ref{mag}) and (\ref{el}) via contraction of a more
general system of relativistic equations for massless fields.

\subsection{Extended Galilei electromagnetism}

In accordance with our analysis of vector field representations of
the Galilei group there exists only one representation, namely,
$D(1,2,1)$ whose carrier space is formed by five-vectors. Such
five-vectors appear naturally in many Galilean models, especially in
those which are constructed via {\it reduction technique} \cite{NN1},
i.e., starting with models invariant with respect to the extended
Poincar\'e group $P(1,4)$  and then reducing them to be invariant w.r.t.
its Galilei subgroup.

As mentioned in our paper \cite{NN1}, there are possibilities to
introduce such different five-component gauge fields which join and
extend the magnetic and electric Galilei limits of the considered relativistic
four-vector-potentials. However, physical meanings of the corresponding
theories have not been clarified. Moreover, as we have seen,
the Maxwell electrodynamics
can be contracted either to magnetic limit (\ref{mag}) or to the
electric limit (\ref{el}), and it has been generally accepted to
think
that it is impossible to formulate a consistent theory which
includes both `electric' and `magnetic' Galilean gauge fields,
(see,
e.g., \cite{santos}).

Accepting  the correctness of this statement, we shall nevertheless show
that in some sense it is possible
to join the `electric' and `magnetic' Galilei gauge fields since the
Galilean five-vector potential appears naturally via contraction of
a relativistic theory. Rather surprisingly, the corresponding
relativistic equations are decoupled to two non-interacting
subsystems whereas their contracted counterparts appear to be
coupled. This is in accordance with the observation presented in
\cite{NN1} that some indecomposable representations of the
homogeneous Galilei group appear via contractions of  completely
reducible representations of the Lorentz group.

Let us begin with relativistic equations for vector-potential
$A^\nu$ \beq\label{relpot}p^\mu p_\mu A^\nu=-ej^\nu\eeq in the
Lorentz gauge, i.e., fulfilling  \beq\label{Lorentz}p_\mu A^\mu=0 \
\ \texttt{or }p_0A^0={\bf p}\cdot{\bf A}.\eeq In equations
(\ref{relpot}) and  (\ref{Lorentz})
$p_\mu=\texttt{i}\frac{\partial}{\partial x^\mu}$, indices $\mu,
\nu$ run over the values $\mu,\nu=0,1,2,3$, $\bf A$ and $\bf p$ are
vector whose components are $A^1, A^2,A^3$ and $p_1,p_2,p_3$
correspondingly.

Let us consider in addition the inhomogeneous d'Alembert equation
for a relativistic scalar field denoted as $A^4$:
\beq\label{scalpot}p^\mu p_\mu A^4=-ej^4.\eeq

Introducing the related vectors of the field strengthes in the
standard form: \beq\label{F}  {\bf H}=\nabla\times{\bf A},\ \ {\bf
E}=\frac{\partial {\bf A}}{\partial x_0}-\nabla A^0, \ {\bf
F}=-\nabla A^4, \ F^0=\frac{\partial A^4}{\partial x_0},\eeq we get
the Maxwell equations for ${\bf E}$ and ${\bf H}$: \beq\label{HH}
\bea{l}\displaystyle\nabla\times {\bf E}-\frac{\partial {\bf
H}}{\partial x_0}=0, \ \ \nabla\cdot{\bf H}=0,\\\\\displaystyle
\nabla\times {\bf H}+\frac{\partial {\bf E}}{\partial x_0}=e{\bf
j},\ \ \ \ \ \ \nabla\cdot{\bf E}=ej^0\eea\eeq and the following
equations for $\bf F$ and $F^0$
\beq\label{FF}\bea{l}\displaystyle\frac{\partial { F^0}}{\partial
x_0}-\nabla\cdot{\bf F}=ej^4,\\\\\displaystyle \nabla\times{\bf
F}=0,\ \ \frac{\partial {\bf F}}{\partial x_0}=\nabla F^0.\eea\eeq

Clearly the system of equations (\ref{HH}) and (\ref{FF}) is
completely decoupled. Its Galilean counterpart obtained using the
In\"on\"u-Wigner contraction appears to be, rather surprisingly,
coupled. This contraction can be made directly for equations
(\ref{HH}) and (\ref{FF}) but we prefer another,
 a more simple way
with potential equations (\ref{relpot}).

The system of equations (\ref{relpot})-(\ref{scalpot}) describes a
decoupled system of relativistic equations for the five-component
function \beq\label{A}A=\texttt{column}(A^1,A^2,A^3,A^0,A^4)=
\texttt{column}({\bf A},A^0,A^4).\eeq Moreover, the components
$(A^1,A^2,A^3,A^0)$ transform as a four-vector and $A^4$ transforms
as a scalar. The related generators (\ref{Smu}) of the Lorentz group
 realize a direct sum
 of representations of the algebra
$so(1,3)$, namely $D(\frac12,\frac12)\oplus D(0,0)$.

In accordance with \cite{NN1} the completely
reduced representation of the Lie algebra of the Lorentz group whose
basis elements are given by equation (\ref{Smu}) can be contracted
either to a direct sum of indecomposable representations of  the
Galilei algebra $hg(1,3)$ or to indecomposable representation
$D(1,2,1)$ of this algebra.

Let us consider the second possibility, i.e., a contraction to the
indecomposable representation. Such contraction is presented in
equations (\ref{Smu})-(\ref{con7}).

Let us demonstrate now that this contraction reduces decoupled
relativistic system (\ref{relpot}) and (\ref{scalpot}) to a system
of the coupled equations invariant with respect to the Galilei
group. Indeed, denoting $U_3A=\texttt{column}({\bf A}',A'^0,A'^4)$
and $U_3j= \texttt{column}({\bf j}'j'^0,j'^4)$ by $ A'$ and $j'$
respectively and taking into account that a non-relativistic
variable $t$ is associated with the relativistic variable $x_0$   by
$x_0=ct\sim\frac1\varepsilon t$, we come to the following system of
equations for the transformed quantities: \beq\label{sys1}{\bf p}^2
A'^k=-e j'^k, \ \ \  \ri \frac{\partial A'^4}{\partial t}={\bf
p}\cdot{\bf A}'.\eeq

Generators of the Galilei group for vectors $A'$ and $j'$ are
expressed in equation (\ref{con7}), so under the Galilei
transformations (\ref{11}) they cotransform in accordance with the
representation $D(1,2,1)$, i.e., \beq\label{trans}A^0\to A^0+{\bf
v}\cdot {\bf A}+\frac{{\bf v}^2}2A^4,\ \ {\bf A}\to{\bf A}+{\bf
v}A^4,\ \ A^4\to A^4,\eeq and
 \beq \label{16}j^4\to j^4, \ \ {\bf j}\to{\bf
j}+{\bf v}j^4,\ j^0\to j^0+{\bf v}\cdot{\bf j}+\frac12v^2j^4.\eeq

Of course, transformations (\ref{11}), (\ref{trans}) and (\ref{16})
keep the system (\ref{sys1}) invariant. In accordance with
(\ref{sys1}) the corresponding field strengthes (compare with
(\ref{F})) \beq\label{RWNB}{\bf W}=\nabla\times{\bf A}',\ \ {\bf
N}=\frac{\partial {\bf A}'}{\partial t}-\nabla  A'^0,\ \ {\bf
R}=\nabla  A'^4, \ \ B=\frac{\partial  A'^4}{\partial t}\eeq satisfy
the following equations \beq\label{coupl}\bea{l}{\cal
C}\equiv\nabla\cdot{\bf N}-\frac{\partial}{\partial
t}B-ej^0=0,\\{\mbox{\boldmath${\cal U}$\unboldmath}} \equiv
\frac{\partial}{\partial t}{\bf R}+\nabla\times{\bf W}-e{\bf
j}=0,\\{\cal A}\equiv\nabla\cdot{\bf
R}-ej^4=0,\\{\mbox{\boldmath${\cal
N}$\unboldmath}}\equiv\frac{\partial}{\partial t}{\bf
W}+\nabla\times{\bf N}=0,\\{\mbox{\boldmath${\cal
W}$\unboldmath}}\equiv\frac{\partial}{\partial t}{\bf R}-\nabla
B=0,\\ {\mbox{\boldmath${\cal R}$\unboldmath}}\equiv -\nabla\times
{\bf R}=0, \texttt{ and}\\ {\cal B}\equiv\nabla\cdot {\bf
W}=0.\eea\eeq

These equations are covariant with
respect to the Galilei group
like (\ref{sys1}). Moreover, the Galilei transformations
for fields ${\bf R}, {\bf W}, {\bf N}$ , $B$ and current $j$ are
given by equations (\ref{fin}) and (\ref{16}) correspondingly.
 In other words, these fields and current $j$ form carrier spaces
of the representations $D(3,1,1)$ and $D(1,2,1)$ of the algebra
$hg(1,3)$, respectively.

In contrast to a decoupled relativistic system of equations
(\ref{HH}) and (\ref{FF}) its Galilei counterpart (\ref{coupl})
appears to be a coupled system of equations for vectors ${\bf R},
{\bf W},\ {\bf N}$ and scalar $B$.

The system of equations equivalent to (\ref{coupl}) was derived in paper
\cite{ijtp} via reduction of generalized Maxwell equations invariant
with respect to the extended Poincar\'e group $P(1,4)$ with one time
and four spatial variables. We have proved that this system is nothing
else than a contracted version of system (\ref{HH}), (\ref{FF})
including the ordinary Maxwell equations and equations for a
four-gradient of the scalar potential. It other words, the system of the
Galilei invariant equations (\ref{coupl}) admits a clear physical
interpretation as a non-relativistic limit of the system of familiar
equations (\ref{HH}) and (\ref{FF}).

\subsection{Reduced Galilean electromagnetism}

In contrast to the relativistic case the Galilei invariant approach
allows to reduce the number of field variables. For example,
considering magnetic limit (\ref{mag}) of the Maxwell equations it
is possible to restrict ourselves to the case ${\bf H}_m=0$ since
this condition is invariant with respect to the Galilei
transformations due to (\ref{tran}). Notice that in a relativistic
theory such condition can be only imposed in a particular frame of
reference and will be violated by the Lorentz transformations.

In the mentioned sense equations (\ref{coupl}) are reducible too.
They are defined on the most extended multiplet of vector and spinor
fields which is a carrier space of an indecomposable representation
of the homogeneous Galilei group. The corresponding representation
$D(3,1,1)$ is indecomposable but reducible, i.e., it includes
subspaces invariant with respect to the Galilei group. This makes
possible to reduce a number of dependent components of equations
(\ref{coupl}) without violating its Galilei invariance.

In this section we shall consider systematically all possible
Galilei invariant constrains which can be imposed on solutions of
the equations (\ref{coupl}) and present the corresponding reduced
versions of the Galilean electromagnetism.

In accordance with (\ref{fin}) and (\ref{16}) vector $\bf R$ and the
fourth component $j^4$ of the current form  invariant subspaces with
respect to the Galilei transformations. Thus we can impose the
Galilei-invariant conditions \beq\label{V=0}{\bf R}=0\ \ \texttt{or
}\nabla A^4=0,  \ \ j^4=0\eeq and reduce system (\ref{coupl}) to the
following one \beq\label{coupl1}\bea{l}\frac{\partial}{\partial
t}\tilde{\bf H}+\nabla\times\tilde{\bf E}=0,\\\nabla\times\tilde{\bf
H}=e{\bf j},\ \nabla\cdot \tilde{\bf H}=0,\\\nabla\cdot\tilde{\bf
E}=\frac{\partial}{\partial t}S+ej^0,\\\nabla S=0,\eea\eeq where we
have used notation $\tilde{\bf H}={\bf W}|_{{\bf R}\equiv0},\
\tilde{\bf E}={\bf N}|_{{\bf R}\equiv0}$ and $S=B|_{{\bf
R}\equiv0}.$

Vectors $\tilde {\bf E},\ \tilde {\bf H}$ and scalar $S$ belong to a
carrier space of the representation $D(2,1,1)$. Their Galilei
transformation laws are \beq\label{9}\bea{l}\tilde {\bf
E}\to\tilde {\bf E}+{\bf v}\times \tilde {\bf H}+{\bf v}S,\ \ \tilde
{\bf H}\to\tilde {\bf H},\ S\to S.\eea\eeq

In accordance with (\ref{9}) $S$ belongs to an invariant subspace of
the Galilei transformations, so we can impose the following additional
 Galilei-invariant
condition \beq\label{10}S=0\ \ \texttt{or }\frac{\partial
A^4}{\partial t}=0. \eeq  As a result we come to
equations (\ref{mag}), i.e., to the magnetic limit of Maxwell's
equations. Thus equations (\ref{mag}) are nothing else than the
system of equations (\ref{coupl}) with the additional Galilei-invariant
constrains (\ref{V=0}) and (\ref{10}).

The Galilei transformations of solutions of equations (\ref{mag})
are given by formulae (\ref{tran}). Again we recognize an invariant
subspace spanned on vectors ${\bf H}_m$, and thus it is possible to
impose the invariant condition \beq \label{17}{\bf H}_m=0 \ \
\texttt{or }{\bf A}=\nabla \varphi,\ \ {\bf j}_m=0,\eeq where
$\varphi$ is a solution of the Laplace equation. As a result we come
to the following system \beq\label{mag1}\bea{l}\nabla\times
\widehat{\bf E}=0, \ \ \nabla\cdot\widehat{\bf E}=e\rho,\eea\eeq
where we have used notations $\widehat{\bf E}={\bf E}_m|_{{\bf
H}_m\equiv0}$ and $\rho=j^0_m|_{{\bf H}_m\equiv0}.$

Equation (\ref{mag1}) remains still Galilei-invariant since both
$\widehat{\bf E}$ and $\rho$ are not changed under the Galilei
transformations. The corresponding potential $\tilde A$ is
constrained by conditions (\ref{V=0}), (\ref{10}) and (\ref{17}).
Moreover, up to gauge transformations it is possible to set ${\bf
A}=0$ in (\ref{17}).

\subsection{Other reductions}

Equations (\ref{coupl1}), (\ref{mag}) and (\ref{mag1}) exhaust all
Galilei-invariant systems which can be obtained starting with
(\ref{coupl}) and imposing additional constraints which reduce the
number of dependent variables. To find the other Galilei-invariant
equations for massless vector fields we shall use observation
 that the Galilei transformations of equations (\ref{coupl}) have the form
 written in relations  (\ref{fin}) if we replace here
 capital letters by calligraphic ones, i.e.,
 ${\bf N}\to {\mbox{\boldmath$\cal N$\unboldmath}},
 {\bf W}\to{\mbox{\boldmath$\cal W$\unboldmath}}, \cdots $.
 Thus the following subsystem of equations (\ref{coupl}) (obtained by excluding
 a self-invariant pair of equations, namely: ${\mbox{\boldmath$\cal N$\unboldmath}}=0$
  and ${\cal C}=0$):

\beq\label{coupl2}\bea{l}{\mbox{\boldmath${\cal U}$\unboldmath}}
\equiv \nabla\times{\bf W}+\frac{\partial}{\partial t}{\bf R}-e{\bf
j}=0,\\{\cal A}\equiv\nabla\cdot{\bf
R}-ej^4=0,\\{\mbox{\boldmath${\cal
W}$\unboldmath}}\equiv\frac{\partial}{\partial t}{\bf R}-\nabla
B=0,\\ -{\mbox{\boldmath${\cal R}$\unboldmath}}\equiv \nabla\times
{\bf R}=0,\\ {\cal B}\equiv\nabla\cdot {\bf W}=0\eea\eeq is
Galilei-covariant too and does not include dependent variables $\bf
N$ and $j^0$. The Galilei transformations of ${\bf W}, {\bf R}, B$ and
${\bf j}, j^4$ remain determined by equations (\ref{fin}) and
(\ref{16}).

Following analogous reasonings it is possible to exclude from
(\ref{coupl2}) the equations
${\mbox{\boldmath$\cal U$\unboldmath}}=0$ and ${\cal B}=0$
and to obtain the system
 \beq\label{coupl4}\bea{l}\nabla\cdot{\bf
R}-ej^4=0,\\\frac{\partial} {\partial t}{\bf R}-\nabla B=0,\\
\nabla\times {\bf R}=0,\eea\eeq which includes only two vector and
two scalar variables. The corresponding potential without loss of
generality reduces to the only variable $A^4$.

The other possibility to reduce system (\ref{coupl2}) consists  to
exclude the equation ${\mbox{\boldmath${\cal W}$\unboldmath}}=0$. As
a result we come to the electric limit for the Maxwell equations
(\ref{el}) for ${\bf W}={\bf H}_e$ and ${\bf R}={\bf E}_e$.

Thus, in addition to (\ref{coupl}), (\ref{coupl1}), (\ref{mag}) and
(\ref{mag1}), we have three other Galilei-invariant systems given by
equations (\ref{coupl2}), (\ref{coupl4}) and (\ref{el}). These
equations admit additional reductions by imposing Galilei-invariant
constrains to their solutions.

Considering (\ref{el}) we easily find that another possible invariant
condition is formed by the pair ${\bf E}_e=0, \ j^4_e=0$ so that
(\ref{el}) reduces
to the following equations \beq\label{111}\nabla\times\hat{\bf
H}=e{\bf j},\ \ \nabla\cdot\hat{\bf H}=0,\eeq where $\hat{\bf H}$
denotes ${\bf H}_e|_{{\bf E}_e\equiv0}$.

Let us return to system (\ref{coupl2}). This system can be reduced
by imposing the Galilei-invariant pair of conditions ${\bf R}=0, j^4=0$ to
the following (decoupled) form: \beq\label{coupl3}\bea{l}\nabla\times\hat{\bf
H}-e{\bf j}=0,\\\nabla\cdot \hat{\bf H}=0,\ \nabla S=0, \eea\eeq
where we have changed the notations ${\bf W}\to\hat{\bf H}$ and
$B\to S$. The corresponding potential reads $A=(A^4,0,{\bf A})$,
where $A^4$ should satisfy the condition $\nabla A^4=0$.

We see that, in contrast to a relativistic theory, there exist a big
variety of equations for massless vector fields invariant with
respect to the Galilei group. The list of such equations is given by
formulae (\ref{mag}), (\ref{el}), (\ref{coupl}), (\ref{coupl1}) and
(\ref{mag1})--(\ref{coupl3}).

\section{ Nonlinear equations for vector fields}

Starting with indecomposable representations of the group $HG(1,3)$
found in \cite{NN1,NN2} it is possible to find out various classes
of partial differential equations invariant w.r.t. the Galilei
group. In the previous sections we have restricted ourselves to
linear Galilean equations for vector and scalar fields and now we
shall present
nonlinear equations. More precisely, we shall study systems of
quasilinear first order equations invariant with respect to the
representations discussed in Section 2.

\subsection{Galilei electromagnetic field in media}

Let us consider the Maxwell equations for electromagnetic field in
a medium \beq\label{B_I}\bea{l}\frac{\partial {\bf
D}}{\partial t}=\nabla\times{\bf H}, \ \ \nabla\cdot{\bf D}=0,
\\\\\frac{\partial {\bf
B}}{\partial t}=-\nabla\times{\bf E}, \ \ \nabla\cdot{\bf
B}=0.\end{array}\eeq Here $\bf E$ and $\bf H$ are vectors of
electric and magnetic field strengths and $\bf D$ and $\bf B$ denote
the corresponding vectors of electric and magnetic inductions. The
system (\ref{B_I}) is underdetermined and has to be completed by
constitutive equations which represent the medium properties. The
simplest constitutive equations correspond to a case where $\bf B$
and $\bf D$ are proportional to $\bf H$ and $\bf E$ respectively,
i.e., \beq\label{C_E}{\bf B}=\mu{\bf H}\ \texttt{  and
 }\ {\bf D}=\kappa {\bf E}.\eeq Here $\mu$ and $\varepsilon$
 are
constants.

In general $\mu$ and $\varepsilon$ can be scalar functions of $\bf
E$ and $\bf H$ so that the related theories are essentially
nonlinear. There are even more complex constitutive equations, e.g.,
\beq\label{C_EE}{\bf B}=\mu{\bf H}+\nu{\bf E},\ \ {\bf D}=\kappa
{\bf E}+\lambda{\bf H},\eeq where $\mu, \nu,\kappa $ and $\lambda$
are some functions of ${\bf H}$ and ${\bf E}$. A popular example of
the constitutive equations is the Born-Infeld system \cite{B_I}
which we consider in the next section.

Let us note that system (\ref{B_I}) by itself, i.e.,  without
constitutive equations, is invariant with respect to a very extended
group which includes both the Poincar\'e and the Galilei groups
as subgroups \cite{NF4}. And just constitutive equations, e.g.,
(\ref{C_E}) or (\ref{C_EE}), reduce this group to the Poincar\'e
group.

Since we are studying Galilean aspects of electrodynamics, it is
naturally to pose a problem, weather there exist such constitutive
equations which reduce the symmetry of system (\ref{B_I}) to the
Galilei group.

For this purpose we shall search for Galilei-invariant constitutive
equations in the form (\ref{C_EE}). Such equations are
Galilei-invariant provided
$ \mu, \nu,\kappa $ and $\lambda$ are invariants of Galilei
transformations and, in addition,
\beq\label{B_ICO}\sigma\kappa=\nu, \ \ \mu=\sigma\lambda,
\eeq
where $\sigma $ is an invariant of the Galilei group.

The Galilei transformations of vectors ${\bf H}$,  ${\bf E}$, ${\bf
D}$ and ${\bf B}$, which keep equations (\ref{B_I}) invariant, have
the form: \beq\label{1111}{\bf E}\to{\bf E}+{\bf v}\times{\bf B},\
{\bf H}\to{\bf H}-{\bf v}\times{\bf D},\ {\bf D}\to{\bf D},\ {\bf
B}\to{\bf B}.\eeq A list of independent bilinear invariants of these
transformations reads: \beq\label{222} {\bf E}\cdot{\bf B},\ {\bf
H}\cdot {\bf D},\ {\bf D}^2,\ {\bf B}^2, \ {\bf E}\cdot{\bf D}-{\bf
H}\cdot{\bf B},\ {\bf B}\cdot{\bf D}.\eeq Notice that all the other
invariants are their functions.

 Thus we have found the general Galilei-invariant equations for an
 electromagnetic
 field in media in the form (\ref{B_I}) with constitutive equations
 (\ref{C_EE}), where $\mu,\nu,\varepsilon$ and  $\lambda$ are arbitrary
 functions of invariants (\ref{222}) satisfying conditions (\ref{B_ICO}). In the
 next section we shall present Galilean versions of the Born-Infeld
 equations.

\subsection{Galilean Born-Infeld equations}

The relativistic Born-Infeld equations include system (\ref{B_I})
and the constitutive equations  \beq\label{B_I1}{\bf D}=\frac1L
({\bf E}+({\bf B}\cdot{\bf E}){\bf B}),\ \ {\bf H}=\frac1L ({\bf
B}-({\bf B}\cdot{\bf E}){\bf E}),\eeq where $L=(1+{\bf B}^2-{\bf
E}^2-{\bf B}\cdot{\bf E})^{1/2}$. Equations (\ref{B_I}) are
Lorentz-invariant. To figure out the corresponding representation of
the Lorentz group explicitly we represent vectors of its carrier
space in the following form
\beq\label{B_I3}\Psi=\texttt{column}({\bf B},{\bf E},{\bf D},{\bf
H}).\eeq Then the associated generators of the Lorentz group are
written as a direct sum of matrices (\ref{repb}), i.e., as:
\beq\label{B_I4}\hat S_{ab}=\left(\bea{cc}S_{ab}&0\\
0&S_{ab}\eea\right),\
S_{0a}=\left(\bea{cc}S_{0a}&0\\
0&S_{0a}\eea\right).\eeq

The In\"on\"u-Wigner contraction can be found by using direct
sums of the contracting matrices (\ref{con4}), i.e., by
\beq\label{B_I5}V_6=\left(\bea{cc}V_4&\bz_{6\times6}\\\bz_{6\times6}&V_5\eea\right)\
\ \texttt{or
}V_7=\left(\bea{cc}V_5&\bz_{6\times6}\\\bz_{6\times6}&V_4\eea\right).\eeq

First let us apply contraction matrix $V_6$ on $\Psi$ defined in
(\ref{B_I3}). Then the vectors in $\Psi$ will be transformed in such
a way that $\Psi\to\Psi'=\texttt{column}({\bf B}',{\bf E}',{\bf
D}',{\bf H}')=\varepsilon V_6\Psi$, with \beq\label{B_I6}{\bf
E}={\bf E}',\ {\bf B}=\varepsilon{\bf B}',\ {\bf D}={\bf D}',\ {\bf
H}=\varepsilon{\bf H}'.\eeq

Substituting (\ref{B_I6}) into (\ref{B_I}) and (\ref{B_I1}), taking
into account that at the same time $\frac{\partial}{\partial
x_0}\to\varepsilon\frac{\partial}{\partial t},\nabla\to\nabla$ and
equating terms with the lowest powers of $\varepsilon$ we come to the
following system \beq\label{B_I7}\bea{l}\frac{\partial {\bf
D}'}{\partial t}=\nabla\times{\bf H}', \ \ \nabla\cdot{\bf D}'=0,
\\ \nabla\times{\bf E}'=0, \ \ \nabla\cdot{\bf
B}'=0\end{array}\eeq with the constitutive equations\beq\label{B_I8}{\bf
D}'=\frac{{\bf E}'}{\sqrt{1-{\bf E}'^2}},\ \ {\bf H}'=\frac{{\bf
B}'}{\sqrt{1-{\bf E}'^2}}-\frac{({\bf B}'\cdot{\bf E}'){\bf
E}'}{\sqrt{1-{\bf E}'^2}}.\eeq

Equations (\ref{B_I7}), (\ref{B_I8}) are Galilei-invariant.
Moreover, under Galilei boosts vectors ${\bf D}', {\bf H}', {\bf
B}'$ and ${\bf E}'$ cotransform as \beq\label{B_It}\bea{l}{\bf
D}'\to{\bf D}',\ \ {\bf H}'\to{\bf H}'+{\bf v}\times{\bf D}',\\
{\bf B}'\to{\bf B}'+{\bf v}\times{\bf E}',\ \ {\bf E}'\to{\bf
E}'.\eea\eeq

Analogously, starting again with (\ref{B_I}) but using the
contraction matrix $V_7$ instead of $V_6$ we obtain the following
system of the Galilei-invariant equations
\beq\label{B_I9}\bea{l}\nabla\times{\bf H}=0, \ \ \nabla\cdot{\bf
D}=0,
\\\frac{\partial {\bf
B}}{\partial t}=-\nabla\times{\bf E}, \ \ \nabla\cdot{\bf
B}=0,\end{array}\eeq which are supplemented with the Galilei-invariant
constitutive equations \beq\label{B_I10}{\bf D}'=\frac{{\bf E}'}{\sqrt{1+{\bf
B}^2}}+\frac{({\bf B}'\cdot{\bf E}'){\bf B}'}{\sqrt{1+{\bf B}^2}},\
\ \ \ {\bf H}'=\frac{{\bf B}'}{\sqrt{1+{\bf B}^2}}.\eeq

The corresponding transformation laws read
\beq\label{B_Itt}\bea{l}{\bf
D}'\to{\bf D}'-{\bf v}\times{\bf H}',\ \ {\bf H}'\to{\bf H}',\\
{\bf B}'\to{\bf B}',\ \ {\bf E}'\to{\bf E}'-{\bf v}\times{\bf
B}'.\eea\eeq
Thus we have seen that there exist two Galilei limits for the
Maxwell equations in various media which are given by equations
(\ref{B_I7}), (\ref{B_I8}) and (\ref{B_I9}), (\ref{B_I10}). Let us
remark that the other mathematically possible Galilean limits of the
Born-Infeld equations (for instance, by direct sums of two contracting
matrices $V_5\oplus V_5$ or $V_4\oplus V_4$) yield trivial
constitutive equations.

\subsection{Quasilinear wave equations and Galilean Chern-Simon models}

In this section we present nonlinear terms (depending on vectors $\bf W$
 $\bf W$, $\bf V$ and scalar $B$ which can be added to
the system (\ref{coupl}) without violating its Galilei-invariance.
We restrict ourselves to bilinear combinations of these vector
and scalar components.

Using Galilean transformation lows (\ref{fin}) one can verify that
the following scalars $\hat j_0$, $\hat j_4$ and vector ${\bf  j}$:
\beq\label{NN1}\bea{l}\hat j^0=\nu {\bf W}\cdot{\bf N}+
\lambda {\bf R }\cdot{\bf W }+\sigma(B^2-{\bf R }\cdot{\bf N })
+\omega{\bf R }^2+\mu B,\\\hat{\bf j }=\nu(B{\bf  W}+{\bf R }
\times {\bf  N})+\sigma({\bf R }\times{\bf  W}+B{\bf  R})+
\mu{\bf  R},\\ \hat j^4=\nu{\bf  R}\cdot{\bf  W}+\sigma{\bf  R}^2
\eea\eeq
(where Greek letters denote arbitrary parameters)
transform as a five-vector from the carrier space of representation
$D(1,2,1)$ of the $HG(1,3)$. In other words, their transformation
laws
are given by relations (\ref{16}) where, however, the "hats"
are absent.

It follows from the above that Galilei-invariance of system
(\ref{coupl}) will not be violated if we add the terms $\hat j^0$,
$\hat{\bf j}$ and $\hat j^4$ to the first, second and third equations
of the system (\ref{coupl}) respectively. In addition, it is possible to
add terms proportional to $\bf N, W, R$ and $B$ to the fourth, fifth, sixth
and seventh equations correspondingly. As a result we obtain
the following system
\beq\label{NL}\bea{l}\frac{\partial}{\partial
t}B-\nabla\cdot{\bf N}+\nu {\bf W}\cdot{\bf N}+
\lambda {\bf R }\cdot{\bf W }+\sigma(B^2-{\bf R }\cdot{\bf N })
+\omega{\bf R }^2+\mu B=ej^0,\\
\frac{\partial {\bf R}}{\partial t}+
\nabla\times{\bf W}+\nu(B{\bf  W}+{\bf R }
\times {\bf  N})+\sigma({\bf R }\times{\bf  W}+B{\bf  R})+
\mu{\bf  R}=e{\bf j},\\\nabla\cdot{\bf R}+\nu{\bf  R}\cdot{\bf  W}+
\sigma{\bf  R}^2=ej^4,\\\frac{\partial}{\partial t}{\bf
W}+\nabla\times{\bf N}+\rho {\bf N}=0,\\\frac{\partial}{\partial t}{\bf R}-\nabla
B+\rho {\bf W}=0,\\  -\nabla\times
{\bf R}+\rho {\bf R}=0, \texttt{ and}\\ \nabla\cdot {\bf W}+\rho B=0.
\eea\eeq

Starting with (\ref{NL}) and making reduction analogous to ones
considered in Sections 4.3 and 4.4 it is easy to find reduced
versions of this system. Let us present here magnetic and electric
limits of equations (\ref{last}):
 \beq\label{LastLast}\bea{l} \nabla\cdot {\bf E}_m+
 \nu{\bf H}_m\cdot {\bf E}_m=ej^0,\ \ \nabla\times {\bf H}_m=e{\bf j},\\\\\frac{\partial {\bf H}_m}{\partial t}-\nabla\times{\bf E}_m=0,\ \ \nabla\cdot{\bf H}_m=0\eea\eeq
 and
 \beq\label{LLL}\bea{l}\frac{\partial {\bf E}_e}{\partial t}+
 \nabla\times{\bf H}_e+\mu{\bf E_e}=e{\bf j},\ \
 \nabla\times {\bf E}_e=0,\\\\ \nabla\cdot{\bf E}_e+\nu{\bf  E}_e\cdot{\bf  H}_e=ej^4,\ \ \nabla\cdot{\bf H}_e=0.\eea\eeq

\subsection{Lagrangian formulation and conserved currents}

  Formula (\ref{NL}) presents the most general Galilei-invariant
quasilinear system which can be obtained from (\ref{coupl}) by
adding linear terms and quadratic non-linearities. Let us consider
in more detail a particular case of system  (\ref{NL}) which
corresponds to the zero values of arbitrary parameters
$\omega,\sigma, \lambda, \mu $ and $\rho$:
 \beq\label{last}\bea{l}\frac{\partial}{\partial
t}B-\nabla\cdot{\bf N}+\nu {\bf W}\cdot{\bf N}
=ej^0,\\\frac{\partial {\bf R}}{\partial t}+ \nabla\times{\bf
W}+\nu(B{\bf  W}+{\bf R } \times {\bf  N})=e{\bf
j},\\\nabla\cdot{\bf R}+\nu{\bf  R}\cdot{\bf
W}=ej^4,\\\frac{\partial}{\partial t}{\bf W}+\nabla\times{\bf
N}=0,\\\frac{\partial}{\partial t}{\bf R}-\nabla B=0,\\
\nabla\times {\bf R}=0, \texttt{ and}\\ \nabla\cdot {\bf W}=0.
\eea\eeq

Let us remark two things:

 First, equations in (\ref{last}) for
$\nu=0$ coincide with equations (\ref{RWNB}) describing "extended
Galilei electromagnetism", see Section 4.2. Thus vectors $\bf N, W,
R$ and scalar $B$ can be expressed via derivatives of five-vector
potential $A=(A^0,{\bf A}, A^4)$ in accordance with (\ref{F}).

Secondly, equations (\ref{last}) admit a Lagrangian formulation. The
corresponding Lagrangian reads:
 \beq\label{lagragra}\bea{l} L=\frac12\left(B^2-{\bf W}^2\right)-
 {\bf N\cdot R}+\nu (A^4{\bf W\cdot N}
 + A^0{\bf R\cdot W}
 -{\bf A}\cdot({\bf W}B+{\bf R}\times{\bf N}))\\\\
 -e(A^4j^0+A^0j^4-{\bf A\cdot j}). \eea \eeq

 Lagrangian (\ref{lagragra}) includes products of potential $A$ and field strengthes and in this aspect can be treated as a Galilean version of the Chern-Simon Lagrangian \cite{Chern}. On the other hand, the system (\ref{last}) is nothing but a Galilei-invariant
analogue of the Carroll-Field-Jackiw model \cite{Jackiw}. This model
was formulated with a view to examine the possibility of Lorentz and
CPT violations in Maxwell's electrodynamics and is invariant neither
w.r.t. the Lorentz nor w.r.t. the Galilei transformations. Equations
(\ref{last}) can be derived via contraction of a generalized CFJ
model which will be shown elsewhere.

Lagrangian (\ref{lagragra}) is invariant with respect to the Galilei
group which includes in particular shifts of time and spatial
variables. Thus in the case $e=0$ we can find the related
energy-momenta tensor whose components are given in the following
equation:
\beq\label{tensor}\bea{l}{T^0}_0=\frac12(B^2+W_bW_b),\\
{T^0}_a=\varepsilon_{abc}N_bW_c-B{N}_a,\\
{T^a}_0=BR_a+\varepsilon_{abc}R_bW_c,\\
{T^a}_b=N_aR_b+N_bR_a-W_aW_b+\delta_{ab}(\frac12(B^2+W_nW_n)-R_nW_n).\eea
 \eeq

Tensors (\ref{tensor}) satisfy the continuity equations \beq
\frac{\partial}{\partial t}{T^\nu}_0+\frac{\partial}{\partial
x_a}{T^a}_0=0,\ \ \nu=0,1,2,3\eeq and so generate conserved
quantities. Moreover, the energy density $\cal E$ and momentum
density $\cal P$ for a system described by equations (\ref{last})
with $e=0$ are associated with ${T^0}_0$ and ${T^0}_0$ and so can be
written in the following form: \beq {\cal E}=\frac12(B^2+{\bf
W}^2),\ \ {\cal P}={\bf N}\times{\bf W}-B{\bf N}.\eeq

It is interesting to note that the energy-momenta tensor
(\ref{tensor}) is valid also for the linear
 version
of system
(\ref{last}), i.e.,  when $e=\nu=0$.  In other words,
 like in (1+2) dimensional Shern-Simon model the "interaction"
 terms with coupling constant $\nu$ do not affect the energy-momenta
 tensor. This fact gives one more argument to specify
 (\ref{lagragra}) as a Galilean Shern-Simon Lagrangian.

\section{Discussion}

The revision of classical results \cite{lebellac} associated with
Galilean electromagnetism done in the present paper appears to be
possible due to our knowledge of indecomposable representations of
the homogeneous Galilei group defined on vector and scalar fields
\cite{NN1}. Thus the present paper completes results of Le Bellac
and L\'evy-Leblond in \cite{lebellac} and presents an extended class
of the Galilei-invariant equations for massless fields. Among them
are decoupled systems of the first order equations which include the
same number of components as the Maxwell equations as well as
equations with other numbers of components. The most extended system
includes ten components while the most reduced one only three.

It is necessary to stress that the majority of the obtained
equations admit clear physical interpretations. For instance
equation (\ref{mag1}) and (\ref{111}) are basic for electro- and
magnetostatics respectively.
   Our procedure of
deducing the Galilei-invariant equations for vector fields used in
the present paper makes interpretations of the equations rather
straightforward, since any obtained equation has its relativistic
counterpart.

We see that a number of the Galilei wave equations for massless
vector fields is rather huge, and so there are many possibilities to
describe an interaction of non-relativistic charged particles with
external gauge fields. Some of these possibilities have been
discussed in \cite{NN1} and \cite{NN3} (se also \cite{FN2},
\cite{NF3} and \cite{ijtp}, \cite{santos}). Starting with the found
equations and using the list of functional invariants for the
Galilean vector fields presented in \cite{NN2} it is easy to
construct nonlinear models invariant with respect to the Galilei
group, including its supersymmetric extensions. Some examples of
non-linear models have been discussed in Section 5. In particular
Galilean versions of Born-Infeld and Chern-Simon systems are
deduced here.

 Let us note that in the case $\nu=e=0$ Lagrangian
 (\ref{lagragra}) can be reduced to the massless field part of the
 Lagrangian
 discussed in paper \cite{Abreu}. Our contribution is a
 demonstration how the related Euler-Lagrange equations (\ref{last})
  can be obtained via contraction of a relativistic system and a short
  discussion of the related conservation laws.
 Moreover, we have presented a much more general  Galilei-invariant
 non-linear system (\ref{NL}) which
 in principle cannot be obtained within
 {\it reduction approach} used in \cite{Abreu}, i.e., starting with
 systems invariant
 w.r.t. the extended Poincar\'e group $P(1,4)$ (group of motions of
 flat (1+4)-dimensional Minkowski space) and then reducing them
  to
Galilei-invariant systems.

 The
main result presented in this paper is a complete description of all
linear first order Galilei invariant equations for massless vector
and scalar fields. Equations which can be obtained via contractions
of relativistic systems are enumerated in Sections 4.2-4.4, the
general description of Galilean first order systems for vector and
scalar fields is given in the Appendix were all covariant differential
forms for such fields are presented. In addition, we present an
 extended class of non-linear Galilean systems.

\section {Appendix. Covariant differential forms}

To complete our analysis of Galilei-invariant linear wave equations for vector and scalar fields we present a full list of first order differential forms which transform as indecomposable vectors sets under the Galilei transformations. In this way we shall in fact describe general linear Galilei-invariant equations of first order for scalar and vector fields.

Using exact transformation laws given by equations (\ref{11}) and (\ref{fin}) it is not difficult to find the corresponding transformations for derivatives of vector fields. Such derivatives can transform as scalars, vectors or second rank tensors under rotations. Restricting ourselves to those forms which transform as vectors or scalars we obtain the following indecomposable sets of them:
\beq\label{dif}\bea{l}\texttt{For } D(0,1,0):\ \ \{{\mbox{\boldmath${\cal R}$\unboldmath}}_1=\nabla A\};\\\texttt{For } D(1,0,0):\ \ \{{\mbox{\boldmath${\cal R}$\unboldmath}}_2=-\nabla\times{\bf R}\}\ \ \texttt{and } \{{\cal{A}}_1=\nabla\cdot{\bf R}\}\\
\texttt{For }D(1,1,0):\ \ \{ {\mbox{\boldmath${\cal R}$\unboldmath}}_2,\ {\mbox{\boldmath${\cal W}$\unboldmath}}_2=\frac{\partial {\bf R}}{\partial t}-\nabla{B}\}\supset\{{\mbox{\boldmath${\cal R}$\unboldmath}}_2\}, \texttt{ and }\{{\cal{A}}_1\};\\
\texttt{For } D(1,1,1):\ \ \{ {\cal B}_1=\frac12\left(\frac{\partial A}{\partial t}+\nabla\cdot{\bf U}\right), \ {\mbox{\boldmath${\cal W}$\unboldmath}}_1=\nabla\times {\bf U},\ {\mbox{\boldmath${\cal R}$\unboldmath}}_1\}\supset(\{{\cal B}_1,{\mbox{\boldmath${\cal R}$\unboldmath}}_1\},\\\ \ \ \ \ \ \ \ \ \ \ \ \ \ \ \ \ \ \ \ \ \ \ \{{\mbox{\boldmath${\cal W}$\unboldmath}}_1,\ {\mbox{\boldmath${\cal R}$\unboldmath}}_1\},\ \{{\mbox{\boldmath${\cal R}$\unboldmath}}_1\}), \   \texttt{ and }\{{\cal{A}}_2=\frac{\partial A}{\partial t}-\nabla\cdot{\bf U}\};\\
\texttt{For } D(2,0,0):\ \ \{ {\mbox{\boldmath${\cal U}$\unboldmath}}_1=\frac{\partial \bf R}{\partial t}+\nabla\times{\bf W},\  {\cal A}_1\}\supset \{{\cal A}_1\}\\\ \ \ \ \ \ \ \ \ \ \ \ \ \ \ \ \ \ \ \ \ \ \texttt{ and } \{{\cal B}_2=\nabla\cdot{\bf W},\  {\mbox{\boldmath${\cal R}$\unboldmath}}_2\}\supset\{{\mbox{\boldmath${\cal R}$\unboldmath}}_2\};\\\texttt{For } D(1,2,1):\ \ \{{\mbox{\boldmath${\cal N}$\unboldmath}}_1=\frac{\partial{\bf U}}{\partial t}-\nabla C,\ {\mbox{\boldmath${\cal W}$\unboldmath}}_1,\ {\mbox{\boldmath${\cal R}$\unboldmath}}_1,\ {\widetilde{\cal B}}_1=\frac{\partial A}{\partial t}\}\supset\{{\mbox{\boldmath${\cal W}$\unboldmath}}_1,\ {\mbox{\boldmath${\cal R}$\unboldmath}}_1,\ {\widetilde{\cal B}}_1=\frac{\partial A}{\partial t}\}\supset\\\ \ \ \ \ \ \ \ \ \ \ \ \ \ \ \ \ \ \ \ \ \ \ (\{\widetilde{\cal B}_1,{\mbox{\boldmath${\cal R}$\unboldmath}}_1\},\{{\mbox{\boldmath${\cal W}$\unboldmath}}_1,\ {\mbox{\boldmath${\cal R}$\unboldmath}}_1\},\ \{{\mbox{\boldmath${\cal R}$\unboldmath}}_1\}), \   \texttt{ and }\{{\cal A}_2\};\\\texttt{For } D(2,1,0):\ \ \{{\mbox{\boldmath${\cal W}$\unboldmath}}_2,\ {\mbox{\boldmath${\cal R}$\unboldmath}}_2,\ {\cal B}_2\}\supset(\{{\cal B}_2,{\mbox{\boldmath${\cal R}$\unboldmath}}_2\},\{{\mbox{\boldmath${\cal W}$\unboldmath}}_2,\ {\mbox{\boldmath${\cal R}$\unboldmath}}_2\},\ \{{\mbox{\boldmath${\cal R}$\unboldmath}}_2\}),\ \texttt{ and }\{{\cal A}_1\};\\\texttt{For } D(2,1,1):\ \ \{{\mbox{\boldmath${\cal K}$\unboldmath}}_1=\frac{\partial {\bf R}}{\partial t}+\nabla\times {\bf K},{\mbox{\boldmath$\widetilde{\cal R}$\unboldmath}}_1=-\nabla A,\ {\cal A}_1\}\supset(\{{\mbox{\boldmath$\widetilde{\cal R}$\unboldmath}}_1\},\ \{{\cal A}_1\})\\\ \ \ \ \ \ \ \ \ \ \ \ \ \ \ \ \ \ \ \ \ \ \texttt{ and      }\ \ \ \ \ \{\widetilde{\cal B}_2=\nabla\cdot{\bf K}-\frac{\partial{A}}{\partial t},\ {\mbox{\boldmath${\cal R}$\unboldmath}}_2\}\supset\{{\mbox{\boldmath${\cal R}$\unboldmath}}_2\};\\\texttt{For } D(2,2,1):\ \ \{{\mbox{\boldmath${\cal K}$\unboldmath}}_1=\frac{\partial {\bf R}}{\partial t}+\nabla\times {\bf K},\ {\mbox{\boldmath$\widetilde{\cal R}$\unboldmath}}_1,\ {\cal A}_1\}\supset(\{{\mbox{\boldmath$\widetilde{\cal R}$\unboldmath}}_1\},\ \{{\cal A}_1\})\\\ \ \ \ \ \ \ \ \ \ \ \ \ \ \ \ \ \ \ \ \ \  \texttt{ and }
\{\widetilde{\cal B}_2,\ {\mbox{\boldmath${\cal W}$\unboldmath}}_2,\ {\mbox{\boldmath${\cal R}$\unboldmath}}_2\}\supset(\{\widetilde{\cal B}_2,{\mbox{\boldmath${\cal R}$\unboldmath}}_2\},\{{\mbox{\boldmath${\cal W}$\unboldmath}}_2,\ {\mbox{\boldmath${\cal R}$\unboldmath}}_2\},\ \{{\mbox{\boldmath${\cal R}$\unboldmath}}_2\}); \\ \texttt{For } D(3,1,1):\ \ \{{\mbox{\boldmath${\cal N}$\unboldmath}}_2=\frac{\partial {\bf W}}{\partial t}+\nabla\times{\bf N},\ {\mbox{\boldmath${\cal W}$\unboldmath}}_2,\ {\mbox{\boldmath${\cal R}$\unboldmath}}_2 , \ {\cal B}_2\}\supset\{{\mbox{\boldmath${\cal W}$\unboldmath}}_2,\ {\mbox{\boldmath${\cal R}$\unboldmath}}_1,\ {{\cal B}}_2=\frac{\partial A}{\partial t}\}\supset\\\ \ \ \ \ \ \ \ \ \ \ \ \ \ \ \ \ \ \ \ \ \ \ (\{{\cal B}_2,{\mbox{\boldmath${\cal R}$\unboldmath}}_2\},\{{\mbox{\boldmath${\cal W}$\unboldmath}}_2,\ {\mbox{\boldmath${\cal R}$\unboldmath}}_2\},\ \{{\mbox{\boldmath${\cal R}$\unboldmath}}_2\}), \\\ \ \ \ \ \ \ \ \ \ \ \ \ \ \ \ \ \ \ \ \ \ \ \texttt{ and }\{{\cal C}_1=\frac{\partial B}{\partial t}-\nabla\cdot {\bf N},\ {\mbox{\boldmath${\cal U}$\unboldmath}}_1,\ {\cal A}_1\}\supset\{{\mbox{\boldmath${\cal U}$\unboldmath}}_1,\ {\cal A}_1\}\supset\{{\cal A}_1\}.
\eea\eeq

Transformation properties of the forms presented in equations (\ref{dif}) are described by relations (\ref{fin}) where capital letters should be replaced by calligraphic ones.
The forms given in brackets are closed w.r.t. Galilei transformations.

Equating differential forms given in (\ref{dif}) to vectors with the same transformation properties or to zero we obtain  systems of linear first order equations for Galilei vector fields.
Thus, starting with representation $D(3,1,1)$, equating ${\mbox{\boldmath${\cal N}$\unboldmath}}_1,\ {\mbox{\boldmath${\cal W}$\unboldmath}}_1,\ {\mbox{\boldmath${\cal R}$\unboldmath}}_1$ and $\cal B$ to zero and $\cal C$, ${\mbox{\boldmath${\cal U}$\unboldmath}}_1 $, $\cal A$ to components of five-current $j^0,{\bf j},j^4$ we obtain the  system (\ref{coupl}).

Notice that there are also tensorial differential forms, namely
\beq\label{LAST}\bea{l}Y_{ab}=\nabla_aR_b+\nabla_bR_a,\ \ L_{ab}=\nabla_aN_b+\nabla_bN_a,\ \    Z^1_{ab}=\nabla_aU_b+\nabla_bU_a,\\R_{ab}=\nabla_aW_b+\nabla_bW_a,\ Z^2_{ab}=\nabla_aK_b+\nabla_bK_a-R_{ab},\ T_{ab}=\nabla_aK_b+\nabla_bK_a\eea\eeq
which transform in a covariant manner under the Galilei transformations provided ${\bf R},\ {\bf U},\ {\bf W},\ {\bf K}$ and $\bf N$ are transformed in accordance with (\ref{fin}). To present invariant sets which include (\ref{LAST}) we need the forms given in (\ref{dif}) and also the following scalar and vector forms:
\beq\label{LASTER}\bea{l}G=\frac{\partial B}{\partial t},\ D=\frac{\partial C}{\partial t},\ {\bf G}=\frac{\partial {\bf W}}{\partial t},\ {\bf F}=\frac{\partial \bf N}{\partial t},\ {\bf P}=\frac{\partial {\bf R}}{\partial t},\ {\bf T}=\frac{\partial \bf K}{\partial T},\\\\{\bf X}=\nabla\times{\bf W},\ {\bf S}= \frac{\partial \bf K}{\partial t}-{\bf G},\ {\bf M}=\frac{\partial \bf R}{\partial t}+\nabla B,\  {\bf J}=\frac{\partial \bf U}{\partial t}+\nabla C.\eea\eeq

The related sets indecomposable w.r.t. the Galilei transformations are enumerated in the following formula:
\beq\label{LAST!}\bea{l}\{Y_{ab}\},\ \{Z^1_{ab},
{\mbox{\boldmath${\cal R}$\unboldmath}}_1\},\ \{R_{ab}, Y_{ab}, {\mbox{\boldmath${\cal R}$\unboldmath}}_2\},\
\{Z^2_{ab}, {\mbox{\boldmath${\cal R}$\unboldmath}}_2\},\ \\\\ \{{\bf M},Y_{ab}\},\ \{{\bf P}, Y_{ab}, {\mbox{\boldmath${\cal R}$\unboldmath}}_2\},\ \{G,{\bf M}, Y_{ab}\},\ \{D, Z^1_{ab},  {\mbox{\boldmath${\cal R}$\unboldmath}}_1,{\bf J}, \widetilde{\cal B}\},\\\\ \{{\bf J},\widetilde{\cal B}_1,{\mbox{\boldmath${\cal R}$\unboldmath}}_1, Z^1_{ab}\},\  \{{\bf G}, R_{ab}, Y_{ab}, {\mbox{\boldmath${\cal R}$\unboldmath}}_2,{\bf P},{\mbox{\boldmath${\cal U}$\unboldmath}}_1\},\   \{{\bf S}, Z^2_{ab}, {\mbox{\boldmath${\cal R}$\unboldmath}}_1, {\mbox{\boldmath${\cal U}$\unboldmath}}_2
-{\mbox{\boldmath${\cal K}$\unboldmath}}_2,
\widetilde {\cal B}_1\},\\\\ \{T_{ab},R_{ab},{\bf X},
{\mbox{\boldmath${\cal R}$\unboldmath}}_1\},\
\{T_{ab},R_{ab},{\bf X},{\mbox{\boldmath${\cal R}$\unboldmath}}_1\}
,\  \{T_{ab},R_{ab},{\bf X},{\bf S},{\mbox{\boldmath${\cal R}
$\unboldmath}}_1,{\mbox{\boldmath${\cal K}$\unboldmath}}_2-
{\bf P},\widetilde{\cal B}\},\\\\ \{Y_{ab}, R_{ab},
L_{ab},{\mbox{\boldmath${\cal R}$\unboldmath}}_2,
{\bf P},{\bf X}\},\ \{Y_{ab}, R_{ab}, L_{ab},
{\mbox{\boldmath${\cal R}$\unboldmath}}_2,
{\bf P},{\bf X},{\bf F}, {\bf M},{\bf G},G\}.\eea\eeq


\begin{thebibliography}{99}

\bibitem{lebellac}
Le Bellac M and L\'evy-Leblond J M \NCB{Galilean
electromagnetism}{1973}{14}{217-33}

\bibitem{Hol} Holland P and Brown H R 2003 The Non-Relativistic Limits
of the Maxwell and Dirac Equations: The Role of Galilean and Gauge
Invariance {\it Studies in History and Philosophy of Science}  {\bf
34} 161-87

\bibitem{NN1} de Montigny M,  Niederle J and Nikitin A G \JPA
{Galilei invariant theories. I. Constructions of indecomposable
finite-dimensional representations of the homogeneous Galilei group:
directly and via contractions}{2006}{39}{1-21}


\bibitem{ijtp} de Montigny M, Khanna F C and Santana A E
\IJTP{Nonrelativistic wave equations with gauge
fields}{2003}{42}{649-71}

\bibitem{santos} Santos E S, de Montigny
M, Khanna F C and Santana A E \JPA{Galilean covariant Lagrangian
models}{2004}{37}{9771-91}

\bibitem{Abreu} Abreu, L M and de Montigny, M \JPA{Galilei
covariant models of bosons coupled to a Chern-Simon gauge field}
{2005}{38}{9877-90}



\bibitem{FN1} Fushchich V I and Nikitin A G
\JPA{Reduction of the representations of the generalised Poincar\'e
algebra by the Galilei algebra}{1980}{13}{2319-30}

\bibitem{FN2} Fushchich W I and Nikitin A G 1994 {\em
Symmetries of Equations of Quantum Mechanics} (New York: Allerton
Press)

\bibitem{contraction} In\"on\"u E and Wigner E P
\PNASUS{On the contraction of groups and their
representations}{1953}{39}{510-24}

\bibitem{NN2} Niederle J and  Nikitin AG 2006 Construction and
classification of indecomposable finite-dimensional representations
of the homogeneous Galilei group {\it Czechoslovak
     Journal of Physics} {\bf 56} 1243-50


\bibitem{NN3} Niederle J and Nikitin A G
{Galilei invariant theories. II.
Wave equations for massive fields} ArXiv: 0707.3286


\bibitem{ll1967} L\'evy-Leblond J M \CMP{Non-relativistic particles
and wave equations}{1967}{6}{286-311}

\bibitem{levyleblond} L\'evy-Leblond J M 1971 Galilei group and
galilean invariance, in {\it Group Theory and Applications} Ed. E.M.
Loebl, Vol. II (New York: Academic) 221-299

\bibitem{levycarroll}
Wightman A S 1959 Relativistic Invariance and Quantum Mechanics (Notes by A. Barut) {\it Nuovo Cimento Suppl}, {\bf 14}, 81-94;
\\Hamermesh M \AP{Galilean invariance and the Schr\"odinger equation}{1960}
{9} {518-521}
\bibitem{B_I} Born M and Infeld L 1934
Foundations of the new field theory {\it Proc. Roy. Soc. A}
{\bf 144} 425-251
\bibitem{NF4} V.I. Fushchich and A.G. Nikitin. Symmetries of Maxwell's
equations. Reidel, Dordrecht, 1987.
\bibitem{NF3} Nikitin A G and
Fuschich W I \TMP{Equations of motion for particles of arbitrary
spin invariant under the Galilei group}{1980}{44}{584-592}
\bibitem{Chern} Chern S S and  Simons J 1974 Characteristic forms
and geometric invariants {\it Annals Math.} {\bf 99} 48–69
\bibitem{Jackiw}S.M. Carrol, J.B. Field and R. Jackiw 1990 Limits on
Lorentz- and parity-violating modifications of electrodynamics. {\it
Phys. Rev D} {\bf 41} 1231-1240




















\end{thebibliography}
\end{document}